\documentclass[conference]{IEEEtran}
\IEEEoverridecommandlockouts
\usepackage{cite}
\usepackage{amsmath,amssymb,amsfonts}
\usepackage{algorithmic}
\usepackage{graphicx}
\usepackage{textcomp}
\usepackage{xcolor}
\def\BibTeX{{\rm B\kern-.05em{\sc i\kern-.025em b}\kern-.08em
    T\kern-.1667em\lower.7ex\hbox{E}\kern-.125emX}}
\begin{document}

\title{A Defensive Framework Against Adversarial Attacks on Machine Learning-Based Network Intrusion Detection Systems}

\author{\IEEEauthorblockN{Benyamin Tafreshian}
\IEEEauthorblockA{\textit{Department of Computer Science} \\
\textit{Boston University}\\
Boston, USA \\
bentaf@bu.edu}
\and
\IEEEauthorblockN{Shengzhi Zhang}
\IEEEauthorblockA{\textit{Department of Computer Science} \\
\textit{Boston University}\\
Boston, USA \\
shengzhi@bu.edu}
}

\maketitle
\begin{center}
    \textbf{This work has been accepted to IEEE AI+ TrustCom 2024. Copyright may be transferred without notice, after which this version may no longer be accessible.}
\end{center}
\thispagestyle{plain}
\pagestyle{plain}
\begin{abstract}
As cyberattacks become increasingly sophisticated, advanced Network Intrusion Detection Systems (NIDS) are critical for modern network security. Traditional signature-based NIDS are inadequate against zero-day and evolving attacks. In response, machine learning (ML)-based NIDS have emerged as promising solutions; however, they are vulnerable to adversarial evasion attacks that subtly manipulate network traffic to bypass detection. To address this vulnerability, we propose a novel defensive framework that enhances the robustness of ML-based NIDS by simultaneously integrating adversarial training, dataset balancing techniques, advanced feature engineering, ensemble learning, and extensive model fine-tuning. We validate our framework using the NSL-KDD and UNSW-NB15 datasets. Experimental results show, on average, a 35\% increase in detection accuracy and a 12.5\% reduction in false positives compared to baseline models, particularly under adversarial conditions. The proposed defense against adversarial attacks significantly advances the practical deployment of robust ML-based NIDS in real-world networks.
\end{abstract}


\section{Introduction}

The increasing volume and sophistication of cyberattacks pose a significant challenge to the modern network security. As organizations increasingly rely on interconnected systems, including IoT technologies, the need for effective cybersecurity measures has become more urgent. Among the various tools used to safeguard these networks, NIDS play a critical role in identifying and mitigating potential malicious activities. Traditionally, these systems have been based on signature detection methods, where predefined rules and patterns are used to identify known attacks. While these approaches are effective for detecting well-documented threats, they struggle with identifying novel or rapidly evolving zero-day attacks—exploits that leverage previously unknown vulnerabilities.

To overcome the limitations of signature-based NIDS, the focus has shifted more towards ML-based anomaly detection. ML-based NIDS analyze network traffic behavior and use statistical models to detect deviations from established norms, which allows them to flag suspicious activity. These systems hold the potential to detect previously unseen attacks by learning the underlying behavior of benign network traffic. However, despite their advantages, ML-based NIDS are extremely vulnerable to adversarial attacks. In such attacks, adversaries manipulate network traffic in subtle ways to deceive the ML models, causing them to misclassify malicious activity as benign. These adversarial attacks have become a growing concern because they exploit weaknesses in the model’s decision boundaries, allowing attackers to bypass detection.

A key aspect of adversarial attacks on NIDS involves manipulating mutable network features—such as packet sizes, network timings, and flow patterns. These features can often be altered by attackers in an attempt to evade detection, as small changes can make malicious traffic appear normal to a ML model. However, in real-world networks, attackers typically do not have full control over certain features like timing or routing behaviors due to network infrastructure constraints and the nature of shared resources. Additionally, mutable features are often interdependent, meaning that changing one feature, such as packet size, may indirectly affect others, like timing or flow patterns, because of the underlying structure of network protocols. These constraints are frequently overlooked in current research, which tends to assume that attackers can manipulate features freely~\cite{b1,b2}.

Another significant gap in the existing research on ML-based NIDS is the lack of robust defense mechanisms to protect against adversarial attacks. Most studies have either focused on improving detection accuracy under non-adversarial conditions \cite{b3, b4, b5, b6} or on generating adversarial attacks to evaluate vulnerabilities in these systems \cite{b7, b8, b9}. This gap underscores the need for stronger, more resilient defenses that can effectively mitigate the risks posed by adversarial threats in real-world network environments.

In this paper, we propose a novel defensive framework to enhance the security and robustness of ML-based NIDS against adversarial attacks by integrating key strategies: adversarial training, dataset balancing, advanced feature engineering, ensemble learning, and extensive model fine-tuning. Adversarial training, which exposes ML models to both normal and adversarial samples, improves their ability to detect manipulated inputs. To address dataset imbalance, we apply tailored oversampling techniques that mitigate bias toward benign traffic. Advanced feature engineering focuses on key traffic features and their interactions, making it harder for attackers to manipulate them without detection. Ensemble learning, combining multiple ML models, strengthens robustness by reducing reliance on any single model. Finally, extensive model fine-tuning optimizes detection accuracy while minimizing false positives and ensuring low-latency performance with GPU-based parallel computation.

To evaluate the effectiveness of our framework, we conducted experiments using two well-established datasets in the field of NIDS: NSL-KDD~\cite{b10} and UNSW-NB15~\cite{b11}. Our experimental results demonstrate that the proposed framework significantly improves detection accuracy while also reducing the rates of false positives and false negatives, especially in the presence of adversarial attacks. This substantial performance gain illustrates the importance of integrating adversarial training, feature engineering, ensemble learning, and model fine-tuning into ML-based NIDS.

\vspace {2pt}\noindent\textbf{Contributions.} The contributions are summarized as follows: 
\vspace {0pt}\noindent$\bullet$ We propose a novel adversarial attack generation methodology that ensures attacks are both realistic and practical. This is achieved by introducing perturbations that comply with network protocols, preserve feature interdependencies, and maintain attack feasibility within real-world network environments. The use of a genetic algorithm ensures that the generated adversarial examples remain both stealthy and effective, making them challenging for ML-based NIDS to detect.
    
\vspace {0pt}\noindent$\bullet$ We introduce a novel defensive framework to improve the resilience of ML-based NIDS. The framework integrates multiple strategies, including adversarial training, protocol-aware feature engineering, balancing techniques, and extensive hyperparameter tuning. By incorporating domain-specific knowledge of network traffic and protocol standards, the framework strengthens the model's ability to detect adversarial manipulations that exploit weaknesses in ML models.
    
\vspace {0pt}\noindent$\bullet$ We test our framework using two widely recognized benchmark datasets, NSL-KDD and UNSW-NB15, to evaluate performance under both regular and adversarial scenarios. Specifically, our framework achieves a 35\% improvement in detection accuracy and reduces the false positive rate by 12.5\% on average compared to baseline models, particularly when adversarial attacks are present.
    
The rest of this paper is organized as follows: Section II explores ML-based NIDS and adversarial attacks. Section III reviews related works in the field. Section IV provides a detailed explanation of how we mount our adversarial attack on ML models, focusing specifically on its impact on ML-based NIDS. Section V details the core defensive strategies used in this work, including adversarial training, dataset balancing, advanced feature engineering, ensemble modeling, and hyperparameter tuning. Section VI presents the experimental results, highlighting performance metrics and insights into model robustness. Finally, Section VII concludes the paper.

\section{Background Information}

\subsection{ML-based NIDS}

ML-based NIDS detect malicious activities by identifying anomalies in network traffic patterns. Unlike traditional signature-based systems, ML-based NIDS use historical data to learn and detect both known and unknown threats, such as zero-day attacks. Key ML models include supervised techniques like Decision Trees, Random Forests, SVMs, and Logistic Regression, as well as unsupervised approaches like k-Means Clustering and Autoencoders. Deep learning models, including CNNs, RNNs, and LSTMs, excel in processing high-dimensional and sequential data, offering improved detection for complex attack patterns and anomalies \cite{b12, b13, b16, b17, b18}.

\subsection{Overview of Adversarial Attacks on ML-based NIDS}

Adversarial attacks on ML-based NIDS involve subtle modifications to input data that mislead the ML model into misclassifying malicious traffic as benign. Formally, adversaries introduce a perturbation $\boldsymbol{\delta}$ to the input feature vector $\mathbf{x}$ to produce $\hat{\mathbf{x}}$, where:

\begin{equation*}
    \hat{\mathbf{x}} = \mathbf{x} + \boldsymbol{\delta}, \quad \text{and} \quad f(\mathbf{x}) \neq f(\hat{\mathbf{x}})
\end{equation*}

The perturbations must meet constraints such as minimal magnitude, feature specificity, and protocol compliance to remain undetected:

\begin{itemize}
    \item \textbf{Minimal Perturbation:} Ensure $\|\boldsymbol{\delta}\|_p \leq \epsilon$ to avoid detection.
    \item \textbf{Feature-Specific Constraints:} Continuous features are perturbed, while categorical features remain immutable.
    \item \textbf{Protocol Compliance:} Modified traffic adheres to standards like TCP, UDP, and ICMP.
\end{itemize}

Adversaries solve an optimization problem to minimize $\|\boldsymbol{\delta}\|_p$ while ensuring $\hat{\mathbf{x}} \in C$, the set of valid inputs adhering to protocol standards.

\subsection{Flows, Packets, and ML-Based Pattern Detection}

Network traffic analysis involves packets aggregated into flows. Packets, the basic transmission units, are analyzed for header anomalies or unauthorized protocol usage. However, flows offer a richer context by tracking sequences of packets with shared attributes, such as IP addresses and transport protocols. This allows detection of more sophisticated attacks like SYN flooding, incomplete handshakes, or session hijacking.

ML models analyze flows to identify deviations in patterns, such as abnormal flow frequencies (indicating DDoS attacks) or flow size variations (suggesting data exfiltration). Models like RNNs and LSTMs are particularly effective for sequential flow analysis, enabling the detection of coordinated or multi-stage attacks. By grouping flows, NIDS can uncover patterns spanning multiple connections, such as botnet activity or port scanning. Behavioral profiling further enhances detection by learning normal communication patterns and flagging deviations.

\begin{table}[h!]
    \centering
    \caption{Comparison of Analysis Levels}
    \renewcommand{\arraystretch}{1.2}
    \begin{tabular}{|c|p{2.75cm}|p{2.75cm}|} 
        \hline
        \textbf{Level of Analysis} & \textbf{Description} & \textbf{Useful For Detecting} \\
        \hline
        Single Packets & Analysis of individual packets, focusing on headers and payload. & Header anomalies, unauthorized protocol use, malformed packets. \\
        \hline
        Flows & Sequence of packets with shared attributes (e.g., source/destination IP, protocol). & SYN flooding, incomplete handshakes, slow-rate DDoS. \\
        \hline
        Groups of Flows & Multiple flows between the same source and destination. & Coordinated attacks, multi-stage intrusions, port scanning. \\
        \hline
    \end{tabular}
    \label{table_analysis_levels}
\end{table}

\section{Related Works}

Adversarial attacks on ML-based NIDS have garnered significant attention due to the increasing reliance on ML to detect sophisticated cyber threats. Recent studies have focused on system vulnerabilities and strategies to mitigate these attacks.

\subsection{Proposed Defensive Mechanisms for ML-based NIDS}

Wang et al. \cite{b19} and Reza et al. \cite{b20} explore robust defenses for ML-based NIDS, such as data augmentation, robustness certification, and protocol-aware feature manipulation, to address adversarial attacks and distribution shifts. These works emphasize the importance of feature engineering and real-world constraints in improving NIDS robustness. Building on these approaches, our framework incorporates adversarial training, balancing techniques, and advanced feature engineering to address distribution shifts and enhance resilience against a broader range of attacks.

\subsection{Adversarial Attack Generation and Evasion Techniques}

Peng et al. \cite{b21} highlight how small perturbations can undermine DoS detection and propose enhancements to boundary-based methods to improve resilience against adversarial inputs. Zhang et al. \cite{b24} investigate Universal Adversarial Perturbations (UAP) on DNN-based NIDS, demonstrating significant attack success rates and underscoring the need for robust defensive architectures. While these studies focus on adversarial example generation and exposing vulnerabilities, our work emphasizes realistic attack generation under real-world constraints, ensuring protocol compliance and feature interdependencies. Additionally, we propose robust defenses to counter these attacks effectively.

\subsection{Attack Transferability Across Machine Learning Models}

Piplai et al. \cite{b23} examine attack transferability using GAN-generated adversarial examples that bypass standard NIDS and suggest ensemble defenses to combat such threats. Expanding on this concept, our framework incorporates ensemble models in the fitness function for attack generation, demonstrating how transferable attacks can be mitigated through integrated defensive strategies. This approach strengthens system robustness and reduces evasion risks across classifiers.

\section{Adversarial Attack Generation Methodology}

Existing methods for generating adversarial examples for ML-based NIDS often overlook real-world constraints, such as protocol compliance, feature mutability, and interdependencies. Many assume attackers can freely manipulate features or test attacks under unrealistic conditions, such as post-feature engineering perturbations or testing on the same classifiers used for training. These limitations often lead to biased results, reducing their practical applicability.

To address these issues, we propose a realistic attack generation framework using a genetic algorithm optimized for feature mutability, interdependence, and network constraints. This approach generates adversarial traffic that evades ML-based NIDS in operational environments. The algorithm employs an ensemble classifier as a black-box baseline, simulating real-world conditions and exploring attack portability across unseen models. The ensemble includes Logistic Regression, SVM, Decision Trees, and Random Forest to ensure robust evaluations. 

\noindent\textbf{Initialization.} The algorithm starts with a population of adversarial instances, each representing perturbed attack examples targeting mutable features (e.g., \texttt{duration}, \texttt{src\_bytes}, \texttt{dst\_bytes}). Perturbations dynamically adapt to feature correlations, ensuring plausible network traffic patterns. For example, increasing \texttt{src\_bytes} proportionally adjusts the permissible range for \texttt{dst\_bytes} to maintain consistency.

\noindent\textbf{Fitness Function.} Adversarial instances are evaluated using a multi-objective fitness function balancing evasion effectiveness and perturbation magnitude:
\begin{equation*}
    f(\hat{x}) = P(\text{benign} \mid \hat{x}) - \lambda \cdot \lVert x - \hat{x} \rVert    
\end{equation*}
Here, $P(\text{benign} \mid \hat{x})$ measures the likelihood of the instance being classified as benign, while $\lambda$ penalizes large deviations from the original input. This approach ensures minimal yet effective perturbations while maintaining realistic traffic patterns.

\noindent\textbf{Selection and Reproduction.} Top-performing individuals are selected via tournament selection and recombined through crossover (80\% probability) to inherit both feature values and correlation-aware perturbation ranges. This process enhances the realism and stealth of adversarial instances by respecting feature interdependencies.

\noindent\textbf{Mutation.} A 1\% mutation rate introduces random changes to features while dynamically adjusting correlated feature ranges. For example, increasing \texttt{src\_bytes} during mutation recalibrates the permissible range for \texttt{dst\_bytes}, ensuring changes remain plausible and undetectable.

\noindent\textbf{Iteration and Termination.} The algorithm iterates through selection, crossover, and mutation for up to 1000 generations or until the fitness function converges. Elitism ensures the top 5\% of adversarial instances are retained across generations. This dynamic perturbation system continuously adapts to feature interdependencies, enabling the generation of adversarial examples that effectively evade detection while mimicking legitimate network traffic.

By integrating correlation-aware perturbations and real-world constraints, our framework produces stealthy and effective adversarial instances. These examples align with network traffic patterns, enhancing their ability to bypass NIDS in operational settings.

\section{Building Robust NIDS against Adversarial Attacks}

Building a resilient NIDS against adversarial attacks requires a multi-layered defense strategy that integrates adversarial training, robust feature engineering, and careful model fine-tuning. Adversarial training strengthens models by exposing them to adversarial examples, improving their ability to detect manipulations. Feature engineering enhances resilience by focusing on high-quality, attack-resistant features, while model fine-tuning ensures classifiers are adaptable and optimized for real-world environments. These combined defenses fortify the system against diverse attack vectors and enhance overall performance.

\subsection{Adversarial Training}

Adversarial training is a widely recognized defense mechanism for mitigating adversarial attacks. This method trains the model on both normal network traffic and adversarial examples designed to deceive the system, enabling it to better identify and classify maliciously altered inputs. By incorporating adversarial samples, the NIDS becomes more resilient to subtle perturbations and manipulation.

However, generating realistic adversarial examples that simulate real-world attacks remains a key challenge. If adversarial samples do not reflect actual attack patterns, the model risks overfitting to synthetic data, impairing its ability to generalize to unseen scenarios. This can reduce effectiveness in detecting both normal and slightly perturbed malicious traffic, thereby compromising robustness.

To overcome these challenges, we incorporated Monte Carlo perturbations during adversarial training, targeting mutable features while maintaining protocol compliance. This approach generated realistic adversarial examples, exposing the model to diverse, real-world scenarios and improving its resilience to evasion attacks. Additionally, we applied the Synthetic Minority Over-sampling Technique (SMOTE) to balance the dataset by oversampling minority classes, ensuring adequate representation of both adversarial and benign traffic. Combining SMOTE with adversarial training enhanced the model’s performance and generalizability, enabling it to detect under-represented attack types without being biased by imbalanced data.

\subsection{Feature Extraction, Selection, and Engineering}

Feature extraction, selection, and engineering are vital for constructing robust ML-based NIDS that can withstand adversarial attacks. These processes enable the system to focus on the most relevant features, reducing the attack surface and improving interpretability. Carefully engineered features are also more resistant to adversarial manipulation, making it harder for attackers to exploit vulnerabilities.

\subsubsection{Feature Extraction and Selection}

Feature extraction transforms raw network traffic data into structured information usable by ML models. Key data points include packet sizes, timing, protocol types, and header fields, with protocol-specific features being particularly critical due to unique vulnerabilities in protocols like TCP, UDP, and ICMP. For example, features like SYN-ACK sequences or retransmission rates in TCP can reveal manipulation attempts.

Feature selection identifies the most informative attributes for the model, such as traffic flow statistics, protocol usage metrics, and anomaly indicators. This step enhances detection efficiency, reduces susceptibility to irrelevant data, and improves generalization across attack types and network environments.

\subsubsection{Feature Engineering for Security}

Feature engineering complements extraction and selection by creating domain-specific features that enhance the model's ability to detect adversarial attacks. A key consideration is distinguishing between mutable features, like packet size and timing, which attackers can alter without breaking protocols, and immutable features, such as certain protocol flags, which are harder to manipulate without triggering alarms or disruptions.

Adversarial perturbations of mutable features are constrained by protocol limitations. Excessive changes to packet sizes or flow rates can disrupt connections, making attacks detectable. In bidirectional protocols like TCP, attackers must manipulate both communication sides without breaking the connection, increasing detection likelihood.

Our approach emphasizes protocol-aware feature engineering, ensuring feature dependencies are respected. For example, perturbing TCP flags must adhere to protocol rules; violations would result in unrealistic traffic patterns. Leveraging protocol-specific knowledge improves detection accuracy and minimizes false positives.

Additionally, feature engineering aggregates traffic data over time windows or multiple segments, helping the model detect subtle anomalies undetectable in single-packet analyses. For instance, tracking statistical properties (e.g., mean, variance) of network flows over time reveals long-term attack patterns, slow-moving anomalies, or multi-stage intrusions. By continuously refining protocol-aware features, the NIDS remains effective across a broad spectrum of adversarial techniques.

\subsection{Model Construction and Ensemble Learning}

For model construction, we employ an ensemble of classifiers, including traditional models such as Logistic Regression and Random Forest, as well as deep learning models like LSTM and MLP. The ensemble approach improves robustness by combining the strengths of multiple models, ensuring that the system is not reliant on any single model’s weaknesses. The fitness of each model is evaluated using both normal and adversarial traffic, ensuring that the NIDS remains effective in real-world scenarios.

\section{Evaluation}

We evaluated our proposed defensive framework using two widely recognized NIDS benchmark datasets: NSL-KDD~\cite{b10} and UNSW-NB15~\cite{b11}. The NSL-KDD dataset, an improved version of KDD Cup 1999, addresses issues like redundant records and data imbalance and includes four attack types: Denial of Service (DoS), Probe, User to Root (U2R), and Remote to Local (R2L). UNSW-NB15 represents modern network traffic with diverse attack types such as Fuzzers and Backdoors. Created using the IXIA PerfectStorm tool, it includes 49 features, providing a comprehensive evaluation benchmark.

Our evaluation assessed the defensive framework across configurations combining strategies such as \textit{Adversarial Training}, \textit{Balancing}, \textit{Feature Preprocessing}, \textit{Feature Engineering}, and \textit{Model Fine-Tuning}. Models were tested on two datasets: \textit{regular testing data} (unaltered traffic) and \textit{adversarial testing data} (perturbed traffic designed to evade detection).

\subsection{Evaluation Metrics}

We evaluated each model using the following metrics:
\begin{itemize}
    \item \textbf{Accuracy}: The proportion of correctly classified instances.
    \begin{equation*}
        \text{Accuracy} = \frac{\text{TP} + \text{TN}}{\text{TP} + \text{TN} + \text{FP} + \text{FN}}
    \end{equation*}
    
    \item \textbf{Precision}: The proportion of true positives out of the predicted positives.
    \begin{equation*}
        \text{Precision} = \frac{\text{TP}}{\text{TP} + \text{FP}}
    \end{equation*}
    
    \item \textbf{Recall}: The proportion of true positives out of the actual positives.
    \begin{equation*}
        \text{Recall} = \frac{\text{TP}}{\text{TP} + \text{FN}}
    \end{equation*}
    
\end{itemize}

\subsection{Classifier Ensembles and Improvement Process}

We evaluate two classifier ensembles: the \textit{Traditional Classifier Ensemble} and the \textit{Deep Learning Ensemble}. The evaluation process is cumulative, beginning with a baseline, followed by the inclusion of additional defensive strategies. The improvement process proceeds as follows:

\begin{itemize}
    \item \textbf{Baseline}: Initial model evaluation with no defensive strategies.
    \item \textbf{Adversarial Training \& Balancing}: Applying adversarial training and balancing (e.g., SMOTE).
    \item \textbf{Feature Engineering \& Preprocessing}: Adding feature engineering and preprocessing techniques.
    \item \textbf{Model Fine-Tuning}: Applying final model fine-tuning for optimal performance.
\end{itemize}

Each classifier ensemble is evaluated on both the \textit{NSL-KDD} and \textit{UNSW-NB15} datasets, with results for both \textit{normal testing datasets} (unaltered traffic) and \textit{adversarial testing datasets} (perturbed traffic).

\subsection{Performance on NSL-KDD and UNSW-NB15 Datasets}

Tables \ref{table_performance_nslkdd} and \ref{table_performance_unsw} summarize the performance of the Traditional Classifier Ensemble (TC) and Deep Learning Ensemble (DC) on the \textit{NSL-KDD} and \textit{UNSW-NB15} datasets. Results are reported for both normal and adversarial testing datasets using metrics such as \textit{Accuracy}, \textit{Precision}, and \textit{Recall} at each stage of the cumulative improvement process.

Configurations with adversarial training consistently improved robustness against adversarial inputs by reducing false negatives, though they introduced a slight increase in false positives. Balancing techniques like SMOTE mitigated this trade-off, particularly in high-dimensional datasets such as \textit{UNSW-NB15}, where false positive rates would have been higher without balancing.

Adversarial training was most effective on adversarial datasets but had limited impact on unaltered datasets. In these cases, SMOTE directly improved recall and reduced missed detections. Preprocessing steps such as scaling and feature interaction further enhanced model performance, while advanced feature engineering improved generalization, especially for complex classifiers.

Fine-tuning provided incremental performance gains, particularly in tree-based and deep learning models, by optimizing performance for nuanced scenarios. Together, these strategies significantly improved accuracy and recall across all evaluated configurations.

\begin{table*}[htpb]
    \caption{Performance on NSL-KDD Dataset}
    \renewcommand{\arraystretch}{1.2}
    \footnotesize
    \begin{tabular}{|p{2.4cm}|c|p{1.25cm}|p{1.2cm}|p{1cm}|p{1.35cm}|p{1.46cm}|p{1.46cm}|} \hline
        \textbf{Classifier Ensemble} & \textbf{Improvement Stage} & \textbf{Normal Accuracy} & \textbf{Normal Precision} & \textbf{Normal Recall} & \textbf{Adversarial Accuracy} & \textbf{Adversarial Precision} & \textbf{Adversarial Recall} \\ \hline
        TC Ensemble & Baseline & 0.713 & 0.887 & 0.588 & 0.396 & 0.399 & 0.050 \\ \hline
        TC Ensemble & + Adversarial Training \& Balancing & 0.865 & 0.915 & 0.832 & 0.742 & 0.792 & 0.644 \\ \hline
        TC Ensemble & + Feature Engineering \& Preprocessing & 0.905 & 0.945 & 0.890 & 0.822 & 0.876 & 0.759 \\ \hline
        TC Ensemble & + Model Fine-Tuning & 0.931 & 0.967 & 0.915 & 0.920 & 0.952 & 0.910 \\ \hline
        DL Ensemble & Baseline & 0.746 & 0.863 & 0.684 & 0.431 & 0.592 & 0.157 \\ \hline
        DL Ensemble & + Adversarial Training \& Balancing & 0.894 & 0.920 & 0.853 & 0.815 & 0.859 & 0.722 \\ \hline
        DL Ensemble & + Feature Engineering \& Preprocessing & 0.931 & 0.946 & 0.901 & 0.872 & 0.903 & 0.797 \\ \hline
        DL Ensemble & + Model Fine-Tuning & 0.960 & 0.953 & 0.982 & 0.940 & 0.933 & 0.969 \\ \hline
    \end{tabular}
    \label{table_performance_nslkdd}
\end{table*}

\begin{table*}[htpb]
    \caption{Performance on UNSW-NB15 Dataset}
    \centering
    \renewcommand{\arraystretch}{1.2}
    \footnotesize
    \begin{tabular}{|p{2.4cm}|c|p{1.25cm}|p{1.2cm}|p{1cm}|p{1.35cm}|p{1.46cm}|p{1.46cm}|} \hline
        \textbf{Classifier Ensemble} & \textbf{Improvement Stage} & \textbf{Normal Accuracy} & \textbf{Normal Precision} & \textbf{Normal Recall} & \textbf{Adversarial Accuracy} & \textbf{Adversarial Precision} & \textbf{Adversarial Recall} \\ \hline
        TC Ensemble & Baseline & 0.750 & 0.703 & 0.957 & 0.254 & 0.137 & 0.064 \\ \hline
        TC Ensemble & + Adversarial Training \& Balancing & 0.860 & 0.813 & 0.972 & 0.684 & 0.579 & 0.512 \\ \hline
        TC Ensemble & + Feature Engineering \& Preprocessing & 0.913 & 0.871 & 0.990 & 0.792 & 0.679 & 0.611 \\ \hline
        TC Ensemble & + Model Fine-Tuning & 0.944 & 0.912 & 0.996 & 0.942 & 0.916 & 0.987 \\ \hline
        DL Ensemble & Baseline & 0.821 & 0.966 & 0.722 & 0.397 & 0.844 & 0.150 \\ \hline
        DL Ensemble & + Adversarial Training \& Balancing & 0.885 & 0.939 & 0.805 & 0.743 & 0.877 & 0.689 \\ \hline
        DL Ensemble & + Feature Engineering \& Preprocessing & 0.932 & 0.967 & 0.893 & 0.874 & 0.909 & 0.786 \\ \hline
        DL Ensemble & + Model Fine-Tuning & 0.977 & 0.993 & 0.969 & 0.963 & 0.981 & 0.956 \\ \hline
    \end{tabular}
    \label{table_performance_unsw}
\end{table*}

\section{Conclusion}

In this paper, we proposed a comprehensive defensive framework to enhance the resilience of machine learning-based NIDS against adversarial attacks by integrating five key defensive strategies: \textit{Adversarial Training}, \textit{Balancing}, \textit{Feature Preprocessing}, \textit{Feature Aggregation/Engineering}, and \textit{Model Fine-Tuning}. A critical aspect of our approach involved generating realistic adversarial samples that adhered to network protocol constraints and feature interdependencies. These samples were not only used to evaluate the framework's robustness but also to refine it through adversarial training, enabling the models to adapt to challenging attack scenarios. This interplay between adversarial sample generation and defense strategies ensured that the framework effectively mitigates evolving threats, making it practical for real-world deployment.

\end{document}